\documentclass[showpacs,preprintnumbers,aps]{revtex4}
\begin{document}
\preprint{USM-TH-188}
\title{Confinement from spontaneous breaking of
scale symmetry}
\author{Patricio Gaete}
\email {patricio.gaete@usm.cl} \affiliation{Departamento de
F\'{\i}sica, Universidad T\'ecnica F. Santa Mar\'{\i}a,
Valpara\'{\i}so, Chile}
\author{Eduardo Guendelman}
\email {guendel@bgu.ac.il} \affiliation{Physics Department, Ben
Gurion University, Beer Sheva 84105, Israel}
\date{\today}

\begin{abstract}
We show that one can obtain naturally the confinement of static
charges from the spontaneous symmetry breaking of scale invariance
in a gauge theory. At the classical level a confining force is
obtained and at the quantum level, using a gauge invariant but
path-dependent variables formalism, the Cornell confining
potential is explicitly obtained. Our procedure answers completely
to the requirements by 't Hooft for ''perturbative confinement''.
\end{abstract}
\maketitle

\section{Introduction}

The question of confinement in gauge theories has been approached
with the use of many different techniques and ideas, like lattice
gauge theory techniques \cite{Wilson} and non-perturbative
solutions of Schwinger-Dyson's equations \cite{Zachariesen}. All
these approaches have the goal of proving the existence of a
linear potential between static quark sources. Of particular
interest to us is the analysis by 't Hooft \cite{'t Hooft} on the
requirements for ''perturbative confinement''. In this work we
will show how is it possible to satisfy 't Hooft's requirements
from a model where the necessary ingredients for confinement arose
from spontaneous symmetry breaking of scale invariance.

The study of the spectrum of heavy quark-antiquark systems is very
well understood. However, as is well known, the binding energy of
an infinitely heavy quark-antiquark pair represents a fundamental
concept in $QCD$ which is expected to play an important role in
the understanding of quark confinement. In this respect we recall
that the famous "Cornell potential" \cite{Eichten} was postulated
in order to simulate the features of $QCD$, that is,
\begin{equation}
V =  - \frac{{\kappa}}{r} + \frac{r}{{a^2 }}, \label{Cornell}
\end{equation}
here $a$ is a constant with the dimensions of length. From a field
theory point of view 't Hooft has shown that confinement can be
associated to the appearance of a linear term in the dielectric
field ${\bf D}$ (that dominates for low $|{\bf D}|$) that appears
in the energy density:
\begin{equation}
U({\bf D}) = \rho _{str} |{\bf D}|, \label{'t Hooft}
\end{equation}
the proportionality constant being the coefficient of the linear
potential, that is, $\rho _{str}  \approx \frac{1}{{a^2 }}$.

It is worthwhile remarking at this point that the appearance of
the scale $a$ in the Cornell potential (\ref{Cornell}) and in the
't Hooft approach is very important. One should take notice that
the original gauge field theory does not have any scales.
Furthermore gauge theories with no scale have a symmetry which is
associated to this, scale invariance. Thus it follows that the
confinement phenomena breaks the scale invariance as the Cornell
potential (\ref{Cornell}) explicitly shows by introducing the
scale $a$.

In this paper we will investigate the connection between scale
symmetry breaking and confinement. In particular we will show the
appearance of the Cornell potential (\ref{Cornell}) as well as the
't Hooft relation (\ref{'t Hooft}) after spontaneous breaking of
scale invariance in a specific model \cite{Guendelman1}. The
quark-antiquark potential is then calculated using the gauge
invariant variables formalism \cite{Gaete1}.

We also draw attention to the fact that the scale invariant model
studied \cite{Guendelman1} introduces, in addition to the standard
gauge fields also maximal rank gauge field strengths of four
indices in four dimensions, $F_{\mu \nu \alpha \beta }  = \partial
_{\left[ \mu  \right.} A_{\left. {\nu \alpha \beta } \right]}$
where $ A_{\nu\alpha\beta} $ is a three index potential. The
integration of the equations of motion of the $ A_{\nu\alpha\beta}
$ field introduces a constant of integration $M$ which breaks the
scale invariance. As we will see, the linear term in the Cornell
potential arises from the constant of integration $M$. When $M=0$
the equations of motion reduce to those of the standard gauge
field theory.

A short note on the history of these kind of models and this way
of breaking scale invariance is in order here. This technique for
breaking scale invariance was used first in generally covariant
theories containing a dilaton field in
Refs.\cite{Guendelman2,Guendelman3}, in the context of a general
type of models which were studied (in non scale invariant form)
before \cite{Guendelman4}. This approach has also been used to
dynamically generate the tension of strings and
branes \cite{Guendelman5}. In
Refs.\cite{Guendelman2,Guendelman3,Guendelman4,Guendelman5} the
maximal rank gauge field strength derives from a potential which
is composite out of $D$-scalars.

In order to calculate the potential energy between a
quark-antiquark pair we will use the gauge-invariant but
path-dependent variables formalism \cite{Gaete1}. Here the
quark-antiquark state is made gauge invariant by the introduction
of a gauge field cloud which is basically the path-ordered
exponential of the gauge field potential along the path where the
two charges are located. This methodology has been used previously
in many examples for studying features of screening and
confinement in gauge theories \cite{Gaete2,Gaete3,Gaete4} and a
preliminary version of this work has appeared before \cite{our preprint}.

\section{Scale invariance breaking and generation of confinement}

We will study the scale symmetry breaking in the context of an
Abelian theory. The non-Abelian generalization presents no
problems \cite{Gaete1}.

Our starting point is the well known action
\begin{equation}
S = \int {d^4 } x\left( { - \frac{1}{4}F_{\mu \nu }^{a} F^{a\mu \nu } }
\right), \label{sib1}
\end{equation}
where $ F_{\mu \nu }^a  = \partial _\mu  A_\nu
^a - \partial _\nu  A_\mu ^a  + gf^{abc} A_\mu ^b A_\nu
^c$. This theory is invariant under the scale symmetry
\begin{equation}
A^{a}_\mu  \left( x \right) \mapsto A_\mu ^ {a\prime}  \left( x \right)
= \lambda A^{a}_\mu  \left( {\lambda x} \right), \label{sib2}
\end{equation}
here $\lambda$ is a constant.

Let us now rewrite (\ref{sib1}) with the use of an auxiliary field
$\omega$
\begin{equation}
S = \int {d^4 } x\left( { - \frac{1}{4}\omega ^2  +
\frac{1}{2}\omega \sqrt { - F_{\mu \nu }^{a} F^{a\mu \nu } } }
\right).\label{sib3}
\end{equation}
From the equation of the $\omega$ field we get
\begin{equation}
\omega  = \sqrt { - F_{\mu \nu }^{a} F^{a\mu \nu } }, \label{sib4}
\end{equation}
and replacing (\ref{sib4}) back into (\ref{sib3}) we get then
(\ref{sib1}). Substituting (\ref{sib4}) in (\ref{sib3}) is a valid
operation because (\ref{sib4}) is a constraint equation. Under a
scale transformation $\omega$ transforms as
\begin{equation}
\omega  \mapsto \lambda ^2 \omega \left( {\lambda x}
\right).\label{sib5}
\end{equation}

Let us now introduce a charge in the theory (\ref{sib3}): we will
now keep the form (\ref{sib3}) but now $\omega$ will not be an
elementary field, rather $\omega$ will be given by
\begin{equation}
\omega  = \varepsilon ^{\mu \nu \alpha \beta } \partial _{\left[
\mu  \right.} A^{a}_{\left. {\nu \alpha \beta } \right]}. \label{sib6}
\end{equation}
Notice that we have introduced a new degree of freedom, the three
index potential and it generates the 4-index field strength
$F^{a}_{\mu \nu \alpha \beta }  \equiv \partial _{\left[ \mu  \right.}
A^{a}_{\left. {\nu \alpha \beta } \right]}$ , a  "maximal rank"  ( of
4-indices in 4-dimensions) field strength. In that case the
equation of motion of $A^{a}_{\nu\alpha\beta}$ is
\begin{equation}
\varepsilon ^{\gamma \delta \alpha \beta } \partial _\beta  \left(
{\omega  - \sqrt { - F^{a\mu \nu } F^{a}_{\mu \nu } } } \right) = 0,
\label{sib7}
\end{equation}
which is integrated to give
\begin{equation}
\omega  = \sqrt { - F^{a}_{\mu \nu } F^{a\mu \nu } }  + M .\label{sib8}
\end{equation}
The integration constant $M$ spontaneously breaks the scale
invariance, since both $\omega$ and $\sqrt { - F^{a\mu \nu }
F^{a}_{\mu \nu } }$ transform as in Eq.(\ref{sib5}) but $M$ does not
transform. Notice that $M$ has the same dimensions as the field
strength $F^{a}_{\mu\nu}$, that is, dimensions of $\left( {length}
\right)^{ - 2}$. We further observe that the variation of the
$A^{a}_\mu$ field produces the following equation
\begin{equation}
\nabla _\mu  \left( {\omega \frac{{F^{a\mu \nu } }}{{\sqrt
{ - F_{\alpha \beta }^b F^{b\alpha \beta } } }}} \right) = \nabla _\mu
  \left[ {\left( {\sqrt { - F_{\alpha \beta }^a F^{a\alpha \beta } }  + M}
   \right)\frac{{F^{a\mu \nu } }}{{\sqrt { - F_{\alpha \beta }^b
   F^{b\alpha \beta } } }}} \right] = 0,
\label{sib9}
\end{equation}
as we will see in the next section, the introduction of the
unusual $M$ term leads to the generation of confinement. Notice
that the term inside the square brackets corresponds to the
''dielectric field $D^{a\mu \nu }$''. One may suspect this because
the consideration of the $M$ term alone is known to lead to such
behavior. In that case the equations of motion are obtained from
an action of the form
\begin{equation}
S = k\int {d^4 x\sqrt { - F_{\mu \nu }^{a} F^{a\mu \nu } } } ,
\label{sib10}
\end{equation}
where $k$ is a constant. Such model leads to confinement, as shown
in Refs. \cite{Aurilia,Amer}, and to string solutions. Among other
properties it is known that electric monopoles do not exist
\cite{Amer}. We will see however that the consideration of the two
terms in (\ref{sib9}) leads to a richer structure in particular to
solutions containing Coulomb and linear parts, as in the Cornell
potential.

\section{Classical solutions and effective actions}

We study now the spherically symmetric case where consider, for
example, an $SU(N)$ gauge theory only with $F^{a}_{0r}\neq0$.
Then by a gauge transformation we can take $F^{a}_{0r}$ in the $1$
direction in color space. This can only be achieved if (up to a
residual Abelian $U(1)$ transformation) the only non vanishing
component of $A^{a}_{\mu}$ is $A^{1o}=A^{0}(r)$. Then the theory
is reduced to an Abelian one and defining $F^{1}_{\mu\nu} \equiv
F_{\mu\nu}$ and taking $F_{0i}=-E_i$ and $F_{ij}=0$, where ${\bf E}=E(r)\hat {\bf
r}$. Then (\ref{sib9}) gives
\begin{equation}
\nabla  \cdot \left( {{\bf E} + \frac{M}{{\sqrt 2 }}\hat {\bf r}}
\right) = 0, \label{sib11}
\end{equation}
which is solved by
\begin{equation}
{\bf E} =  - \frac{M}{{\sqrt 2 }}\hat {\bf r} + \frac{q}{{r^2
}}\hat{\bf r}. \label{sib12}
\end{equation}
The scalar potential $V$ that gives rise to such electric field is
\begin{equation}
V  =  - \frac{M}{{\sqrt 2 }}r + \frac{q}{r}, \label{sib13}
\end{equation}
which is indeed resembles very much the Cornell potential
(\ref{Cornell}). Notice that so far (\ref{sib13}) refers to the
field of one charge and not yet to the interaction energy between
two charges. We will see that such interaction energy also has the
Cornell form, even at the quantum level. Since Abelian solutions
are solutions of the non-Abelian theory, these solutions are also
relevant for the non-Abelian generalization.

Before approaching the quantum theory (which will be treated in
some approximations) we want to define effective actions that give
the equations of motion (\ref{sib9}). Indeed one can easily see
that
\begin{equation}
{\cal L}_{eff}  =  - \frac{1}{4}F^{a}_{\mu \nu } F^{a\mu \nu }  -
\frac{M}{2}\sqrt { - F^{a}_{\mu \nu } F^{a\mu \nu } }, \label{sib14}
\end{equation}
reproduces Eqs. (\ref{sib9}).

Since the full treatment of the quantum theory is rather
difficult, instead of using (\ref{sib14}) we restrict ourselves to
a "truncated" phase space model where we consider spherical
coordinates $(r,\theta,\varphi)$ in addition to time, but where we
set $F_{ij}=0=F_{0\varphi}=F_{0\theta}$ and consider only $(t,r)$
dependence of $F_{0r}$. Then instead of (\ref{sib14}), we consider
(in the Abelian reduction discussed before)
\begin{equation}
S = 4\pi \int {dr} r^2 {\cal L}_{eff}, \label{sib15}
\end{equation}
where
\begin{equation}
{\cal L}_{eff}  = \frac{1}{2}\left( {F_{0r} } \right)^2  -
\frac{{M\sqrt 2 }}{2}F_{0r}. \label{sib16}
\end{equation}
Similar kind of "reduced phase space" which take into account only
the spherical degrees of freedom have been used elsewhere in other
examples, see for example Ref.\cite{Benguria}.

\section{Interaction energy}

As already mentioned, our aim now is to calculate the interaction
energy between external probe sources in the model (\ref{sib15}).
To do this, we will compute the expectation value of the energy
operator $H$ in the physical state $\left| \Phi  \right\rangle$,
which we will denote by $\left\langle H \right\rangle _\Phi$. The
starting point is the two-dimensional space-time Lagrangian
(\ref{sib15}):
\begin{equation}
{\cal L} = 4\pi r^2 \left\{ { - \frac{1}{4}F_{\mu \nu } F^{\mu \nu
} - \frac{{M\sqrt 2 }}{4}\varepsilon _{\mu \nu } F^{\mu \nu } }
\right\} - A_0 J^0, \label{poten1}
\end{equation}
where $J^0$ is the external current. A notation remark, in
(\ref{poten1}), $\mu,\nu=0,1$, also, $ x^1  \equiv r \equiv x$ and
$\varepsilon^{01}=1$.

We now proceed to obtain the Hamiltonian. For this we restrict our
attention to the Hamiltonian framework of this theory. The
canonical momenta read $\Pi ^\mu   =  - 4\pi x^2 \left( {F^{0\mu }
+ \frac{{M\sqrt 2 }}{2}\varepsilon ^{0\mu } } \right)$, which
results in the usual primary constraint $\Pi^0=0$, and $\Pi ^i = -
4\pi x^2 \left( {F^{0i}  + \frac{{M\sqrt 2 }}{2}\varepsilon ^{0i}
} \right)$. The canonical Hamiltonian following from the above
Lagrangian is:
\begin{equation}
H_C  = \int {dx} \left( {\Pi _1 \partial ^1 A^0  - \frac{1}{{8\pi
x^2 }}\Pi _1 \Pi ^1  - \frac{{M\sqrt 2 }}{2}\varepsilon ^{01} \Pi
_1  + A_0 J^0 } \right). \label{poten2}
\end{equation}
Notice that the field ${\raise0.7ex\hbox{${\Pi ^1 }$} \!\mathord{\left/
 {\vphantom {{\Pi ^1 } {4\pi x^2 }}}\right.\kern-\nulldelimiterspace}
\!\lower0.7ex\hbox{${4\pi x^2 }$}}$ plays the role of the ($r$-component)
dielectric field ${\bf D}$ indeed. The term linear in $M$ and ${\Pi ^1 }$
is the term linear in $|{\bf D}|$ that 't Hooft has shown is needed in
order to obtain confinement and which must dominate in the limit $|{\bf D}|\to0$
(here ${\Pi ^1 }\to0$).

The consistency condition ${\dot \Pi _0}=0$ leads to the secondary
constraint $\Gamma _1 \left( x \right) \equiv \partial _1 \Pi ^1 -
J^0=0$. It is straightforward to check that there are no further
constraints in the theory, and that the above constraints are
first class. The extended Hamiltonian that generates translations
in time then reads $H = H_C  + \int d x \left( {c_0 (x)\Pi_0 (x) +
c_1 (x)\Gamma _1 (x)} \right)$, where $c_0(x)$ and $c_1(x)$ are
the Lagrange multipliers. Moreover, it follows from this
Hamiltonian that $ \dot{A}_0 \left( x \right) = \left[ {A_0 \left(
x \right),H} \right] = c_0 \left( x \right)$, which is an
arbitrary function. Since $\Pi_0 = 0$, neither $A^0$ nor $\Pi^0$
are of interest in describing the system and may be discarded from
the theory. The Hamiltonian then takes the form
\begin{equation}
H = \int {dx} \left( { - \frac{1}{{8\pi x^2 }}\Pi _1 \Pi ^1  -
\frac{{M\sqrt 2 }}{2}\varepsilon ^{01} \Pi _1  + c^ \prime  \left(
{\partial _1 \Pi ^1  - J^0 } \right)} \right), \label{poten3}
\end{equation}
where $c^ \prime  \left( x \right) = c_1 \left( x \right) - A_0
\left( x \right)$.

According to the usual procedure we introduce a supplementary
condition on the vector potential such that the full set of
constraints becomes second class. A convenient choice is found to
be \cite{Gaete1,Gaete2,Gaete3,Gaete4}
\begin{equation}
\Gamma _2 \left( x \right) \equiv \int\limits_{C_{\xi x} } {dz^\nu
} A_\nu \left( z \right) \equiv \int\limits_0^1 {d\lambda x^1 }
A_1 \left( {\lambda x} \right) = 0, \label{poten4}
\end{equation}
where  $\lambda$ $(0\leq \lambda\leq1)$ is the parameter
describing the spacelike straight path $ x^1  = \xi ^1  + \lambda
\left( {x - \xi } \right)^1 $, and $ \xi $ is a fixed point
(reference point). There is no essential loss of generality if we
restrict our considerations to $ \xi ^1=0 $. In this case, the
only nontrivial Dirac bracket is
\begin{equation}
\left\{ {A_1 \left( x \right),\Pi ^1 \left( y \right)} \right\}^ *
= \delta ^{\left( 1 \right)} \left( {x - y} \right) -
\partial _1^x \int\limits_0^1 {d\lambda x^1 } \delta ^{\left( 1
\right)} \left( {\lambda x - y} \right). \label{poten5}
\end{equation}

We are now equipped to compute the interaction energy between
pointlike sources in the model (\ref{sib15}), where a fermion is
localized at the origin $ {\bf 0}$ and an antifermion at $ {\bf
y}$. As we have already mentioned, we will calculate the
expectation value of the energy operator $H$ in the physical state
$ |\Phi\rangle$. From our above discussion, we see that
$\left\langle H \right\rangle _\Phi$ reads
\begin{equation}
\left\langle H \right\rangle _\Phi   = \left\langle \Phi
\right|\int {dx} \left( { - \frac{1}{{8\pi x^2 }}\Pi _1 \Pi ^1  -
\frac{{M\sqrt 2 }}{2}\varepsilon ^{01} \Pi _1 } \right)\left| \Phi
\right\rangle . \label{poten6}
\end{equation}
Next, as remarked by Dirac\cite{Dirac}, the physical state can be
written as
\begin{equation}
\left| \Phi  \right\rangle  \equiv \left| {\overline \Psi  \left(
\bf y \right)\Psi \left( \bf 0 \right)} \right\rangle  = \overline
\psi \left( \bf y \right)\exp \left( {ie\int\limits_{\bf 0}^{\bf
y} {dz^i } A_i \left( z \right)} \right)\psi \left(\bf 0
\right)\left| 0 \right\rangle, \label{poten7}
\end{equation}
where $\left| 0 \right\rangle$ is the physical vacuum state. As we
have already indicated, the line integral appearing in the above
expression is along a spacelike path starting at $\bf 0$ and
ending $\bf y$, on a fixed time slice.

Taking into account the above Hamiltonian structure, we observe
that
\begin{equation}
\Pi _1 \left( x \right)\left| {\overline \Psi  \left( y
\right)\Psi \left( 0 \right)} \right\rangle  = \overline \Psi
\left( y \right)\Psi \left( 0 \right)\Pi _1 \left( x \right)\left|
0 \right\rangle  - e\int_0^y {dz_1 } \delta ^{\left( 1 \right)}
\left( {z_1  - x} \right)\left| \Phi  \right\rangle.
\label{poten8}
\end{equation}
Inserting this back into (\ref{poten6}), we get
\begin{equation}
\left\langle H \right\rangle _\Phi   = \left\langle H
\right\rangle _0  + \frac{{e^2 }}{{8\pi }}\int {dx} \frac{1}{{x^2
}}\left( {\int_0^y {dz_1 \delta ^{\left( 1 \right)} } \left( {z_1
- x} \right)} \right)^2  + \frac{{M\sqrt 2 e}}{4}\int {dx} \left(
{\int_0^y {dz_1 } \delta ^{\left( 1 \right)} \left( {z_1  - x}
\right)} \right), \label{poten9}
\end{equation}
where $\left\langle H \right\rangle _0  = \left\langle 0
\right|H\left| 0 \right\rangle$. We further note that
\begin{equation}
\frac{{e^2 }}{2}\int {dx} \left( {\int_0^y {dz\delta ^1 \left(
{z_1  - x} \right)} } \right)^2  = \frac{{e^2 }}{2}L ,
\label{pato}
\end{equation}
with $|y|\equiv L$. Inserting this into Eq.(\ref{poten9}), the
interaction energy in the presence of the static charges will be
given by
\begin{equation}
V =  - \frac{{e^2 }}{{8\pi }}\frac{1}{L} + \frac{{M\sqrt 2
e}}{2}L, \label{poten10}
\end{equation}
which has the Cornell form. In this way the static interaction
between fermions arises only because of the requirement that the
$\left| {\overline \Psi \Psi } \right\rangle$ states be gauge
invariant. Notice that confinement is obtained at a finite value
of the strong coupling just as claimed by 't Hooft \cite{'t
Hooft}.

\section{Conclussions}

We have found that in the context of a model where scale
invariance is spontaneously broken, the Cornell confining
potential between quark-antiquark naturally appears. The solutions
appear also relevant to the non-Abelian generalizations of the
model. Once again, the gauge-invariant formalism has been very
economical in order to obtain the interaction energy, this time
showing a confining effect in $(3+1)$ dimensions. The model
satisfies indeed the 't Hooft basic criterion for achieving
confinement even with finite coupling constant, and in this case
the necessary term for the dependence of the energy density for
low dielectric field (linear in $|{\bf D}|$) discussed by 't Hooft
is here obtained as a result of spontaneous breaking of scale
invariance which introduces the constant $M$. Other aspects of
$QCD$ concern gluon confinement, in addition to the
quark-antiquark confinement we have studied so far. Indeed,
preliminary studies indicate that Eq.(\ref{sib9}), do not support
plane wave solutions, which is a clear hint of gluon confinement.
We will report on these issues in a future publication. In a
separate paper we have studied the effect of both $\sqrt { -
F_{\mu \nu } F^{\mu \nu } }$ and mass terms for gluons \cite{our
PLB} showing that the interplay of both terms leads to a Coulomb
interaction.

\section{ACKNOWLEDGMENTS}

One of us (E. G.) wants to thank the Physics Department of the
Universidad T\'{e}cnica F. Santa Mar\'{\i}a for hospitality. P. G.
was partially supported by FONDECYT (Chile) grant 1050546.

\end{document}